\documentclass[amsmath,amssymb,prb,unsortedaddress,twocolumn]{revtex4-2}
\usepackage{graphicx}
\usepackage{dcolumn}
\usepackage{bm}
\usepackage{color}

\usepackage[normalem]{ulem}

\begin{document}
\title{Magnetic ground states of CrPS$_4$ and  NiPS$_3$ monolayers from long-range exchange interactions}
\author{Balázs Nagyfalusi$^{1}$}
\email{Contact author: nagyfalusibalazs@uniovi.es}
\author{Ritesh Das$^2$}
\author{Alvaro Bermejillo-Seco$^2$}
\author{Herre S. J. van der Zant$^2$}
\author{Yaroslav M.  Blanter$^2$}
\author{Amador Garc\'ia-Fuente$^{1,3}$}
\author{Jaime Ferrer$^{1,3}$}
\email{Contact author: ferrer@uniovi.es}
\affiliation{$^1$Departamento de Física, Universidad de Oviedo, 33007 Oviedo, Spain}
\affiliation{$^2$Kavli Institute of Nanoscience, Delft University of Technology, Lorentzweg 1, 
             2628 CJ, Delft, The Netherlands}
\affiliation{$^3$Centro de Investigación en Nanomateriales y Nanotecnología, CSIC, 33940 El Entrego, Spain}
\date{\today}

\begin{abstract}
We investigate the magnetic properties of monolayer CrPS$_4$ and NiPS$_3$ by combining first-principles calculations, second-principles spin models, and Monte Carlo simulations. 
Unlike conventional approaches that truncate exchange interactions after only a few  shells and determine them by fitting total energies, we extract the magnetic exchange tensors directly from density functional theory using the LKAG formalism and include interactions until numerical convergence is achieved. 
We show that long-range exchange interactions qualitatively modify the magnetic behavior of both materials. 
In CrPS$_4$, they destabilize the previously predicted ferromagnetic ground state and stabilize a spin-spiral phase, reducing the critical temperature to about 21\,K, in agreement with available experiments.
The resulting magnetic phase diagram contains multiple collinear and non-collinear phases that can be tuned by temperature and external magnetic fields. 
In NiPS$_3$, the experimentally observed zigzag antiferromagnetic order only emerges when exchange interactions up to the fifth shell are included. 
These results demonstrate that quantitatively predictive spin models for thiophosphate monolayers require long-range exchange interactions and provide a predictive framework for accurately describing two-dimensional van der Waals magnets.
\end{abstract}

\maketitle

\section{Introduction}
The discovery of intrinsic ferromagnetic order in two-dimensional (2D) CrI$_3$~\cite{Huang2017} has sparked widespread interest in 2D magnetism over the past decade. 
Since then, several van der Waals magnetic compounds have been exfoliated down to a few tens of nanometers and, in some cases, to the monolayer limit~\cite{Gong2017,Fei2018,yang2021,roadmap_2025}. 
In parallel, considerable theoretical effort has been devoted to constructing effective spin Hamiltonians to describe these materials.
Such models are commonly truncated after the first few shells, typically up to third-nearest neighbors see Ref.~\cite{orlando2026} and references therein.
While this approximation is often sufficient, it may fail in systems where competing longer-range exchange interactions qualitatively modify the magnetic ground state.

In this work, we focus on two magnetic 2D materials, NiPS$_3$ and CrPS$_4$, for which we find long-range exchange interactions play a crucial role in determining the magnetic order of their monolayer forms. 
Although both compounds are transition-metal thiophosphate semiconductors, they exhibit markedly different bulk crystallographic and magnetic properties. 
Bulk NiPS$_3$ crystallizes in a hexagonal structure. Each NiPS$_3$ layer displays an antiferromagnetic (AFM) zigzag configuration with spins oriented along an in-plane easy axis, while adjacent layers are coupled ferromagnetically~\cite{joy1992,Wildes2022,lancon2018}. 
In contrast, bulk CrPS$_4$ has a monoclinic lattice. 
Each CrPS$_4$ layer exhibits ferromagnetic (FM) order with spins aligned along the out-of-plane direction, while neighboring layers are coupled antiferromagnetically~\cite{Zhuang2016,Susilo2020}. 
The N\'eel temperatures are 155\,K for NiPS$_3$~\cite{joy1992} and 36\,K for CrPS$_4$~\cite{Louisy1978}.

Upon exfoliation to the monolayer limit, however, the magnetic properties of these materials become less clear.
For NiPS$_3$, theoretical studies generally predict that the bulk zigzag structure persists down to the monolayer limit (see Ref.~\cite{orlando2026} and references therein), whereas experiments indicate a suppression of long-range magnetic order~\cite{kim2019}, attributed to the small magnetic anisotropy.
For monolayer CrPS$_4$, magneto-optic Kerr microscopy experiments report a critical temperature of approximately 23\,K~\cite{son2021,Hou2024} together with out-of-plane ferromagnetic ordering under an applied magnetic field. 
Most theoretical studies, however, predict a ferromagnetic ground state with a substantially higher critical temperature~\cite{Deng2021,Bo2023}, suggesting that important physical ingredients may be be missing from the simplified spin models commonly employed for these systems.

CrPS$_4$ and NiPS$_3$ therefore provide ideal model systems for assessing the predictive power of microscopic spin Hamiltonians. 
Their relatively simple crystal structures, good exfoliability, and air stability~\cite{houmes2024highly,kuo2016exfoliation,Hou2024,wu2023,Wang2022} make them particularly well suited for systematic comparisons between theory and experiment. 
At the same time, their relatively low critical temperatures and sensitivity to competing exchange interactions make them excellent platforms for investigating the role of long-range magnetic interactions in two dimensions.

In this work, we employ first-principles calculations to investigate the electronic and magnetic properties of monolayer CrPS$_4$ and NiPS$_3$. 
The magnetic exchange and anisotropy tensors are determined using our in-house code {\sc grogu}~\cite{Grogu2023}, which implements the Liechtenstein--Katsnelson--Antropov--Gubanov (LKAG) formalism~\cite{Liechtenstein1987}. 
Based on these interactions, we perform Monte Carlo simulations to determine the magnetic ground states, critical temperatures, and magnetic phase diagrams, paying particular attention to the convergence of these properties with the number of exchange shells included in the spin Hamiltonian. 
We also compute the spin-wave spectra using {\sc magnopy}~\cite{Magnopy1,Magnopy2}. 
We show that the widely used three-shell approximation is insufficient to describe the magnetic properties of these thiophosphate monolayers.
Long-range exchange interactions stabilize a spin-spiral ground state in CrPS$_4$ and are essential to recover the experimentally observed zigzag order in NiPS$_3$, demonstrating that predictive spin models for these materials require exchange interactions extending well beyond the conventional short-range approximation.

The article is organized as follows. 
In Sec.~\ref{sec:methods}, we summarize the computational methodology. Sections~\ref{sec:CrPS_results} and~\ref{sec:NiPS_results} present the results for CrPS$_4$ and NiPS$_3$, respectively. 
Finally, Sec.~\ref{sec:conc} summarizes our conclusions.

\section{Methods}
\label{sec:methods}

The electronic structure of the monolayers was calculated using the {\sc siesta} density functional theory~(DFT) code 
where the spin-orbit interaction was included self-consistently~\cite{Soler2002,Cuadrado2012}. 
In-house norm-conserving pseudopotentials  were generated for Cr, Ni, P and S, where the lattice constant, magnetic moment and electronic 
bands were fitted to those obtained from the plane-wave, all-electrons code ELK~\cite{elk}.
A triple-$\zeta$ polarized (TZP) basis set was employed. 
The exchange-correlation potential was treated within the generalized gradient approximation~(GGA) using 
the Perdew-Burke-Ernzerhof~(PBE) functional~\cite{Perdew1996}.
The nature and size of the electronic gap and the main 
features of the magnon spectrum were tested using a range of Hubbard $U$ parameters within the LDA+U 
scheme~\cite{Dudarev1998}. 

After careful convergence and accuracy testing, a  $30 \times 30 \times 1$ Monkhorst-Pack $k$-point grid and a real-space mesh cutoff of up-to 4000\,Ry were adopted to ensure a reliable self-consistent convergence.
Structural optimizations were performed using five to ten initial spin arrangements, that included the experimentally expected 
magnetic ground states (namely out-of-plane FM order for CrPS$_4$  and  in-plane  zigzag AFM order for NiPS$_3$). 
These calculations also included the spin–orbit coupling self-consistently to allow for 
spin-lattice relaxations~\cite{Cuadrado2012}.  
The convergence 
thresholds used in the calculations are summarized in Tbl.~\ref{tab:convergence}.

\begin{table}[h]
    \centering
     \caption{Convergence criteria for the different compounds}
    \label{tab:convergence}
    \begin{tabular}{c|cccc}
              & Energy [eV] &Dens. matr. &Force [eV/\AA]& Stress  [GPa] \\ \hline
       CrPS$_4$ &$3\cdot10^{-5}$& $5\cdot10^{-6}$&$10^{-4}$  & $10^{-4}$\    \\
        NiPS$_3$&$3.3\cdot10^{-5}$&$4\cdot10^{-6}$ &$10^{-3}$  & $10^{-3}$\    
    \end{tabular}
   
\end{table}

The DFT Kohn-Sham Hamiltonian was mapped to the following classical Heisenberg model using the {\sc grogu} code~\cite{Grogu2023}:
\begin{align}
     H_0&=
       \frac{1}{2}\sum_{i,j} \mathbf{e}_i \mathbf{J}^\mathrm{}_{ij} \mathbf{e}_j 
      + \frac{1}{2}\sum_{i,j}  \mathbf{D}_{ij}\cdot\left(\mathbf{e}_i \times\mathbf{e}_j\right) 
      + \sum_{i} \mathbf{e}_i \mathbf{A}_{i} \mathbf{e}_i\,. 
    \label{eq:heis_mod}
\end{align}
Here, $\mathbf{J}_{ij}$ is the symmetric part of the full tensorial $3\times3$ exchange interaction, and $\mathbf{D}_{ij}$ is 
the Dzyaloshinskii-Moriya~(DM) interaction  between magnetic sites $i$ and $j$, whereas $\mathbf{A}_{i}$ is the single-ion magnetic 
anisotropy at site~$i$.
Finally, $\mathbf{e}_i$ refers to a unit vector along the spin direction at site~$i$.  
The leading exchange interactions in both CrPS$_4$ and NiPS$_3$ are the isotropic couplings, here defined
as $J_{ij}=\frac{1}{3}\sum_\alpha{J}_{ij}^{\alpha\alpha} $\,, where  $\alpha=x,y,z$. 
We also define the diagonal exchange anisotropy for a given $ij$ pair as 
$d_{ij}^{\alpha\alpha}=J_{ij}^{\alpha\alpha}-J_{ij}^{zz}$ for $\alpha=x,y$, and for the whole system as $d_{}^{\alpha\alpha}=\sum_{ij}d_{ij}^{\alpha\alpha}$. 
We find that off-diagonal matrix elements of the
symmetric tensor $\mathbf{J}_{ij}$ are typically at or below the $\mu$eV range, so we skip discussing them in what follows.

To assess the convergence of the spin Hamiltonian with interaction range, exchange tensors were progressively truncated after increasing numbers of shells. 
Throughout the manuscript, the isotropic interaction corresponding to the $N$th shell is denoted by $J_N$. 
The convergence of the magnetic ground state and thermodynamic properties with respect to the number of retained shells is analyzed in Sec.~\ref{sec:CrPS_results} and  Sec.~\ref{sec:NiPS_results} for CrPS$_4$ and NiPS$_3$, respectively.
To achieve the required accuracy in these second-principles {\sc grogu} calculations, several k-grids and energy meshes were tested 
and a Monkhorst-Pack $k$-point grid of $30 \times 30 \times 1$  and 100 energy points was found to lead to converged results.

The magnetic ground state and the corresponding critical temperatures were determined using Monte Carlo~(MC) simulations. 
An initial supercell grid of $64\times64$ unit cells was employed, with periodic boundary conditions applied in all directions. 
The systems were annealed from the high-temperature paramagnetic phase in temperature steps of $0.25\,$K. 
The thermal averages at each temperature were calculated from 40 samples each after $5000$ MC step to thermalize the system. For 
the non-collinear structures and near the transition temperature these values were increased to 100 samples and 10000 steps.
The critical temperature was identified from the peak of the heat-capacity curve. 
An external magnetic field $\mathbf{B}$ oriented along the $z$-axis was incorporated into the spin Hamiltonian 
in Eq.~\eqref{eq:heis_mod} via the Zeeman term:
\begin{align}
     H&=H_0   -\mu \sum_i \mathbf{e}_i \mathbf{B}\,, 
    \label{eq:heis_mod_with_B}
\end{align}
where $\mu$ is the magnetic moment obtained in the DFT calculation. 

The {\sc Magnopy}~\cite{Magnopy1,Magnopy2} code was employed to verify the relative stability of the magnetic configurations obtained from Monte Carlo simulations and to compute the spin-wave spectra of both materials.

For both materials, we tested different values of $U$ in the range from 0 to 4\,eV to identify the value providing the best overall agreement with the available experimental data. We followed the same procedure in both cases. First, we selected the Hubbard $U$ by requiring the calculated magnetic interactions to reproduce the shape and order of magnitude of the experimentally measured bulk magnon spectrum. Second, we used the resulting critical temperature as an additional consistency check.

\section{CrPS$_4$ Results}
\label{sec:CrPS_results}
\begin{figure}
    \centering
    \includegraphics[width=0.95\linewidth]{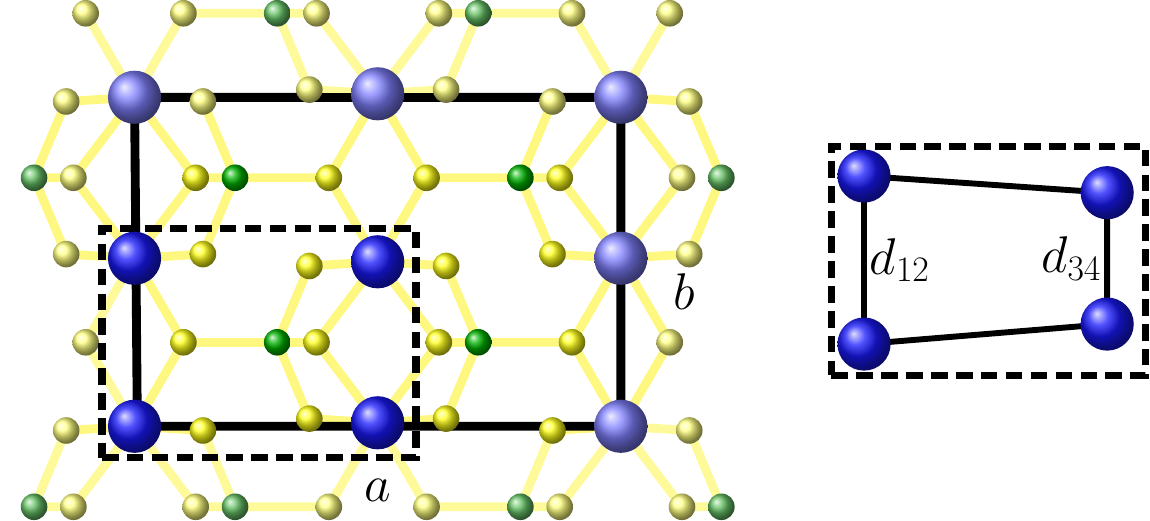} 
\caption{Optimized crystal structure of a CrPS$_4$ monolayer. Blue, green, and yellow spheres denote Cr, P, and S atoms, respectively. The yellow lines indicate the chemical bonds. The lattice constants are $a=10.91$\,\AA\ and $b=7.36$\,\AA, while the inequivalent Cr--Cr bond lengths are $d_{12}=3.75$\,\AA\ and $d_{34}=3.60$\,\AA, as indicated by the black lines.}
    \label{fig:CrPS4_structure}
\end{figure}

CrPS$_4$ crystallizes in a monoclinic lattice in which each Cr atom is coordinated by six S atoms forming a distorted octahedron, as shown in Fig.~\ref{fig:CrPS4_structure}. 
Along the $b$ direction (the $y$ axis), neighboring Cr atoms are connected through shared S atoms, forming one-dimensional chains. 
These chains are linked along the $a$ direction (the $x$ axis) by P atoms, which connect only every second Cr atom. 
This bonding pattern breaks the translational symmetry and gives rise to a weak Cr dimerization along the $b$ direction, as indicated in Fig.~\ref{fig:CrPS4_structure} and recently observed experimentally~\cite{Multian2025}. 
To capture this structural distortion, we employed a unit cell containing four Cr atoms, four P atoms, and sixteen S atoms.

The combined structural and magnetic optimizations were performed for Hubbard parameters in the range $U=0$--$2\,$eV. 
Among these values, $U=0$ yields the best agreement with the experimentally observed magnetic transition around $23\,\mathrm{K}$ reported from MOKE measurements~\cite{son2021} and is therefore adopted throughout this section. 
Within the four-Cr-atom unit cell, the lowest-energy configuration is an out-of-plane ferromagnet with spins aligned along the $z$ direction. 
Ferromagnetic configurations with spins oriented along the $x$ and $y$ directions are only $0.27$ and $0.29$\,meV higher in energy, respectively, reflecting the weak magnetic anisotropy. 
Among the antiferromagnetic configurations considered, including chessboard and stripy arrangements along the $a$ and $b$ directions, the lowest-energy state lies approximately $25\,$meV above the FM-$z$ configuration.
We emphasize that these structural and magnetic relaxations were performed using the four-Cr-atom unit cell shown in Fig.~\ref{fig:CrPS4_structure}, which cannot accommodate the long-wavelength spin-spiral state identified below. 
Consequently, the ferromagnetic configuration is the lowest-energy state accessible within the periodicity imposed during the DFT optimization.

The optimized ferromagnetic structure has lattice constants $a=10.91$\,\AA\ and $b=7.36$\,\AA, with Cr--Cr bond lengths of $d_{12}=3.75$\,\AA\ and $d_{34}=3.60$\,\AA, corresponding to a bond-length difference of approximately $2\,\%$ arising from the structural dimerization.  
The magnetic moment of each Cr atom is $2.95\,\mu_\mathrm{B}$, in good agreement with the experimental bulk value~\cite{Calder2020}. 
The calculated electronic structure exhibits a direct band gap of $0.72\,$eV at the $\Gamma$ point, smaller than the $\sim1.4\,$eV direct gap reported experimentally for bulk CrPS$_4$~\cite{Louisy1978,lee2017}.
 
\begin{figure}
    \centering
     \includegraphics[width=0.7\linewidth]{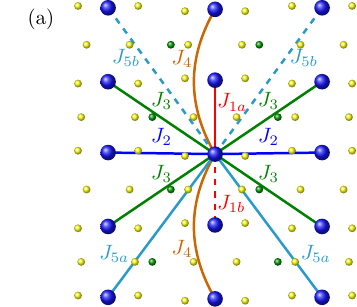}
     \includegraphics[width=0.9\linewidth]{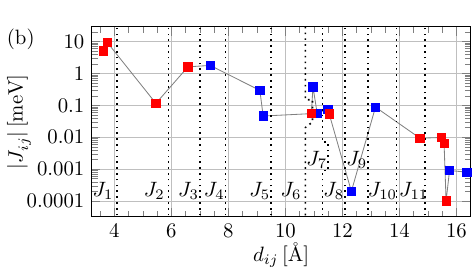}
     \caption{(a) First five shells defining the isotropic exchange interactions $J_N$ in a CrPS$_4$ monolayer. 
     (b) Calculated isotropic exchange interactions $J_N$ as a function of the Cr--Cr distance $d_N$.  Blue and red squares denote antiferromagnetic ($J_N>0$) and ferromagnetic ($J_N<0$) interactions, respectively.
     The dotted lines show the boundaries of exchange shells, marked by $J_N.$}
    \label{fig:CrPS4_J_iso}
\end{figure}

The origin of the magnetic ground state can be understood from the calculated exchange interactions. 
The first five shells are illustrated in Fig.~\ref{fig:CrPS4_J_iso}\,(a). 
Owing to the structural dimerization, the first-neighbor exchange interactions split into two inequivalent couplings, denoted by $J_{1a}$ and $J_{1b}$. 
Previous analyses of neutron-scattering data for bulk CrPS$_4$ have typically employed Heisenberg models including interactions up to the third shell~\cite{Calder2020}, 
and most first-principles studies have extracted similar short-range models by fitting total energies of selected magnetic configurations~\cite{Joe2017,Deng2021}.
Here, the exchange tensors are obtained directly within the LKAG formalism, allowing the interactions for each shell to be determined independently, without requiring a fit to a predefined spin Hamiltonian.

The calculated magnetic interactions are highly spin isotropic. 
Both the single-ion anisotropy and the anisotropic exchange are of the order of only a few $\mu$eV, with $A^{xx}=0.026\,$meV, $A^{yy}=0.035\,$meV, $d^{xx}=-0.008\,$meV, and $d^{yy}=-0.043\,$meV. 
In the following, we therefore focus on the isotropic exchange interactions~$J_{ij}$.

The isotropic exchange interactions are shown in Fig.~\ref{fig:CrPS4_J_iso}\,(b) as a function of the Cr--Cr distance.
The first three shells are all ferromagnetic, in agreement with previous first-principles studies~\cite{Deng2021,Bo2023}. 
Truncating the spin Hamiltonian at this level therefore naturally leads to a ferromagnetic ground state. 
A qualitatively different picture emerges when longer-range interactions are included. 
The fourth-shell interaction, $J_4$, is antiferromagnetic, in agreement with Ref.~\cite{Bo2023}, and its magnitude reaches approximately one quarter of the average first-neighbor coupling.

The importance of $J_4$ lies not only in its magnitude, but also in its geometry. 
As shown in Fig.~\ref{fig:CrPS4_J_iso}\,(a), this interaction couples second neighbors along the $b$ direction and therefore competes directly with the ferromagnetic first-neighbor interactions $J_{1a}$ and $J_{1b}$. 
This competition frustrates the magnetic order and favors a continuous rotation of the spins instead of a collinear ferromagnetic arrangement.
The remaining exchange interactions are predominantly antiferromagnetic and at least one order of magnitude weaker than $J_1$, but they further reinforce this frustrated magnetic state. 
In particular, $J_7$ is an antiferromagnetic interaction coupling next-nearest neighbors along the $b$ direction and, as we show below, provides the additional competing interaction required to destabilize the ferromagnetic ground state.
The combined effect of these long-range interactions stabilizes a spin-spiral ground state propagating along the $b$ direction.

The importance of $J_4$ lies not only in its magnitude, but also in its geometry.
As shown in Fig.~2(a), this interaction couples second-neighbor Cr atoms along the $b$ direction and therefore competes directly with the ferromagnetic first-neighbor interactions $J_{1a}$ and $J_{1b}$.
This competition frustrates the magnetic order and favors a continuous rotation of the spins instead of a collinear ferromagnetic arrangement. 
The remaining exchange interactions are predominantly antiferromagnetic and at least one order of magnitude weaker than $J_1$, but they further reinforce this frustrated magnetic state.

The combined effect of these long-range interactions stabilizes a spin-spiral ground state propagating along the $b$ direction.

\begin{figure}
    \centering
    \includegraphics[width=0.99\linewidth]{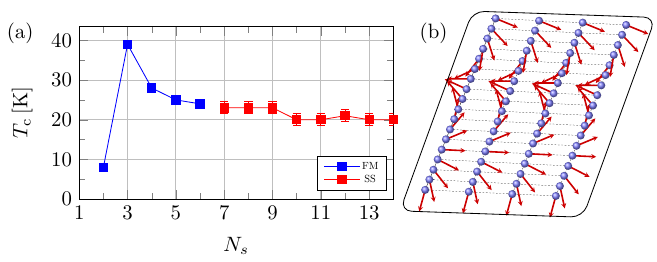}
    \caption{(a) Critical temperature $T_\mathrm{c}$ of a CrPS$_4$ monolayer as a function of the number of exchange shells $N_s$ included in the spin Hamiltonian. 
    Blue and red squares correspond to  ferromagnetic (FM) and spin-spiral (SS) ground states, respectively. 
    The transition from blue to red marks the change in the magnetic ground state upon extending the interaction range. Error bars reflect the uncertainty in determining the specific-heat peak (see Fig.~\ref{fig:CrPS4_susc}(a)).
    (b) Visualization of SS ground state.}
    \label{fig:CrPS4_Tc}
\end{figure}

The consequences of this exchange frustration become apparent in the Monte Carlo simulations. 
Fig.~\ref{fig:CrPS4_Tc}\,(a) shows the evolution of the magnetic ground state and critical temperature as progressively more shells are included in the spin Hamiltonian. 
Although the nearest-neighbor interaction $J_1$ is the dominant exchange coupling, it couples spins along the $b$ direction, whereas the corresponding first-neighbor interaction along the $a$ direction, $J_2$, is comparatively weak. 
Consequently, the magnetic ordering depends sensitively on longer-range exchange interactions.

Restricting the Hamiltonian to the first three shells yields a FM ground state with a critical temperature of approximately $39\,\mathrm{K}$, in agreement with previous theoretical studies~\cite{Deng2021,Bo2023}. 
As additional exchange shells are incorporated, the FM state is progressively destabilized. 
The magnetic ground state converges to a SS configuration (see Fig.~\ref{fig:CrPS4_Tc}\,(b)) once interactions up to the seventh shell are included. 
In contrast, the critical temperature continues to decrease and converges only when exchange interactions up to the tenth shell are retained, yielding $T_\mathrm{c}\approx21\,\mathrm{K}$.
This reduction reflects the increased magnetic frustration introduced by the competing long-range interactions, which weaken the stability of the ordered phase despite leaving the dominant exchange couplings essentially unchanged. 
The resulting transition temperature of 21\,K is in excellent agreement with the experimentally reported value for monolayer CrPS$_4$~\cite{son2021}.
These results demonstrate that reproducing the correct magnetic ground state is not sufficient to accurately describe the finite-temperature behavior; quantitatively predictive spin models require a substantially longer interaction range.

Monte Carlo snapshots reveal that the spin spiral propagates along the $b$ direction (see Fig.~\ref{fig:CrPS4_Tc}\,(b)), consistent with the competing exchange interactions discussed above, and has a wavelength of approximately $6.9\,b$. 
To accurately capture this long-period modulation, the simulations were repeated using supercells extending up to 512 unit cells along the $b$ direction, ensuring negligible finite-size effects. 
The spin configuration is well described by the unit vector
$\mathbf{e}_i=\left(\cos(q\,y_i),0,\sin(q\,y_i)\right)$
where $(x_i,y_i)$ denotes the position of Cr atom $i$ and the spiral wave vector is $q=0.124\,\mathrm{\AA}^{-1}$.

The structural dimerization plays a key role in stabilizing the spin-spiral ground state. 
To quantify its influence, we continuously interpolated between the \emph{ab initio} first-neighbor exchange interactions and the fully symmetric case, where the two inequivalent couplings become identical. 
The interpolated exchange constants were defined as
\begin{align}
J_{1a/b}(\alpha) = \alpha J_{1a/b}^0 + (1-\alpha) J_{1b/a}^0 ,
\label{eq:javerage}
\end{align}
where $\alpha$ varies from 1 (the \emph{ab initio} values) to 0.5 (fully averaged couplings), and $J_{1a/b}^0$ are the calculated, \emph{ab initio} exchange interactions.

As the dimerization is progressively reduced, the wavelength of the spin spiral increases continuously. 
For $\alpha\lesssim0.6$, extended ferromagnetic domains emerge and progressively grow at the expense of the spiral. 
In the fully symmetric limit ($\alpha=0.5$), the spin spiral is replaced by two large ferromagnetic domains separated by a single domain wall. 
These results demonstrate that the structural dimerization enhances the exchange frustration responsible for the spin-spiral state. 
As the dimerization is reduced, this frustration weakens, causing the spiral to unwind and the system to evolve toward collinear ferromagnetic alignment.

\begin{figure}
    \centering
    \includegraphics[width=1\linewidth]{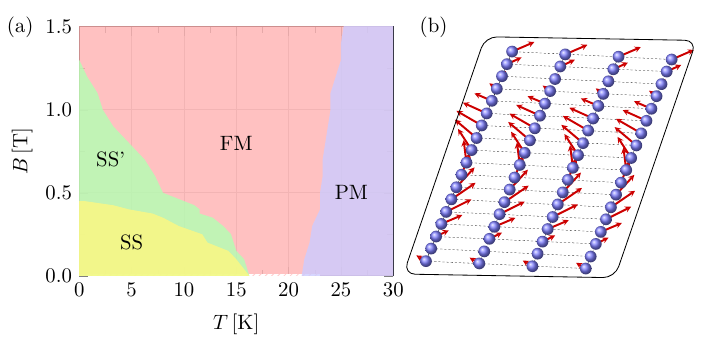}
    \caption{(a) Magnetic phase diagram of a CrPS$_4$ monolayer as a function of temperature and out-of-plane magnetic field. 
    The yellow region corresponds to the spin-spiral (SS) phase (see Fig. \ref{fig:CrPS4_Tc}\,(b)), with spins rotating in the $x$--$z$ plane. 
    The green region denotes the canted spin-spiral (SS') phase (see Fig.\,(b)), in which the spins rotate in the $x$--$y$ plane while developing a finite homogeneous magnetization along the $z$ direction. 
    The red and blue regions correspond to the ferromagnetic (FM) 
    and paramagnetic (PM) phases, respectively.
    (b) Visualisation of SS' ground state.}  

    \label{fig:CrPS4_phase_diag}
\end{figure}

\begin{figure}
    \centering
    \includegraphics[width=0.9\linewidth]{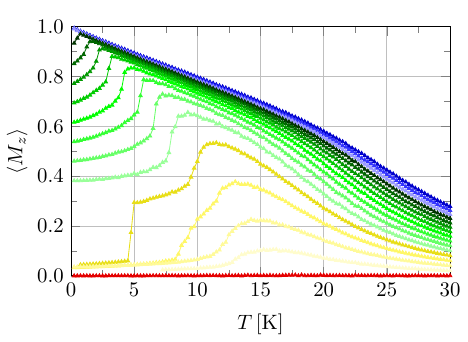}
    \caption{Average out-of-plane magnetization $M_z$ of a CrPS$_4$ monolayer as a function of temperature for magnetic fields between 0 and 1.5\,T in steps of 0.1\,T. 
    The curves are colored according to the sequence of magnetic phases encountered upon cooling: FM (blue), SS'$\rightarrow$FM (green), and SS$\rightarrow$SS'$\rightarrow$FM (yellow). 
    The zero-field ($B=0$) curve is highlighted in red.}
    \label{fig:CrPS4_M}
\end{figure}

We have mapped the $(B,T)$ phase diagram using Monte Carlo simulations, as shown in Fig.~\ref{fig:CrPS4_phase_diag}. 
The corresponding average out-of-plane magnetization, $M_z$, for different values of the external magnetic field is displayed in Fig.~\ref{fig:CrPS4_M}. 
In the absence of an external magnetic field, the average magnetization vanishes and the system remains in the SS phase below the paramagnetic transition. 
Applying a finite magnetic field gives rise to three distinct regimes, which can be classified according to their zero-temperature magnetization.

For magnetic fields below $0.45\,\mathrm{T}$, the average magnetization remains zero and the system forms the SS phase discussed above, in which the spins rotate within the $x$--$z$ plane. 
Between $0.45\,\mathrm{T}$ and $1.3\,\mathrm{T}$, the magnetization becomes finite while remaining smaller than unity. 
Inspection of the Monte Carlo snapshots reveals that the spiral continuously rotates toward the $x$--$y$ plane while developing a finite homogeneous component along the $z$ direction. 
We denote this phase by SS', and describe it by the unit vector
$
\mathbf{e}_i=\left(\sqrt{1-M_z^2}\cos(q,y_i),\sqrt{1-M_z^2}\sin(q,y_i),M_z\right),
$
where $q$ is the spiral wave vector determined above. 
Finally, for magnetic fields exceeding $1.3\,\mathrm{T}$, the spiral is completely suppressed and the system undergoes a transition into a fully polarized ferromagnetic state.

Interestingly, the same sequence of ordered phases also appears as a function of temperature. 
For finite magnetic fields below $1.3\,\mathrm{T}$, the system first orders into the out-of-plane ferromagnetic phase upon cooling from the paramagnetic state. 
Upon further cooling, the ferromagnetic order becomes unstable due to the competing long-range exchange interactions, giving way first to the SS' phase and eventually to the zero-field SS phase. 
As a result, the transition temperature increases gradually with magnetic field, from approximately $21\,\mathrm{K}$ at zero field to about $25\,\mathrm{K}$ at $1.5\,\mathrm{T}$.

\begin{figure}
    \centering
       \includegraphics[width=0.9\linewidth]{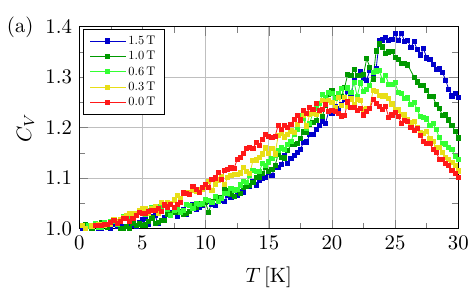}
    \includegraphics[width=0.9\linewidth]{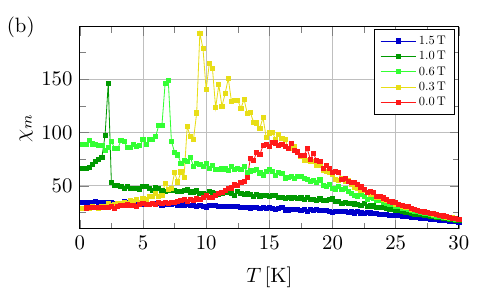}
  \caption{(a) Specific heat $C_V$ and (b) total magnetic susceptibility  $\chi_m$ (Eq. \eqref{eq:susc}) of a CrPS$_4$ monolayer as functions of temperature for different applied magnetic fields. These thermodynamic quantities are used to determine the phase boundaries shown in Fig.~\ref{fig:CrPS4_phase_diag}.}
    \label{fig:CrPS4_susc}
\end{figure}

The phase boundaries were determined from complementary thermodynamic observables. 
The transition to the paramagnetic phase was extracted from the peak of the specific-heat curves shown in Fig.~\ref{fig:CrPS4_susc}\,(a). 
The onset of the ferromagnetic phase was identified from the temperature dependence of the average magnetization in Fig.~\ref{fig:CrPS4_M}, while the boundary between the SS and SS' phases was obtained from the total magnetic susceptibility defined as
\begin{align}
\chi_m =
\frac{1}{k_\mathrm{B}T}
\left(
\langle M\rangle^2 -
\left\langle M_x^2+M_y^2+M_z^2\right\rangle
\right),
\label{eq:susc}
\end{align}
shown in Fig.~\ref{fig:CrPS4_susc}\,(b). 
Note that the system undergoes multiple magnetic phase transitions upon cooling, depending on the applied magnetic field. 
Consequently, features in the heat capacity and magnetic susceptibility can correspond to distinct phase transitions between different magnetic phases. 
For example, in the PM--FM--SS' sequence upon cooling, the high-temperature peak in the heat capacity in Fig.~\ref{fig:CrPS4_susc}\,(a) identifies the transition from the paramagnetic to the ferromagnetic phase. The lower-temperature feature in the magnetic susceptibility in Fig.~\ref{fig:CrPS4_susc}\,(b), in contrast, corresponds to the transition from the FM to the SS' phase.

The measurement of Ref.~\cite{son2021}, which reported a finite out-of-plane magnetization below $23\,\mathrm{K}$ under an applied magnetic field of $0.25\,\mathrm{T}$, is qualitatively consistent with our phase diagram. 
At this field, our simulations predict a ferromagnetic phase above approximately $15\,\mathrm{K}$, followed by an SS$'$ phase between approximately $12$ and $15\,\mathrm{K}$, which carries a finite net out-of-plane magnetization. 
Below approximately $12\,\mathrm{K}$, the equilibrium state becomes the SS phase with vanishing net magnetization. 
The finite out-of-plane magnetization observed experimentally even at $10\,\mathrm{K}$ may reflect a shift of the SS$'$--SS boundary arising from sample-dependent effects such as finite size, strain, or defects, as well as from the uncertainty in the conversion between the applied magnetic field and the Zeeman energy. 
In our simulations, the Zeeman coupling is determined using the magnetic moment obtained from the {\sc SIESTA} calculations, and a difference between this calculated moment and the effective magnetic moment of the experimental system could lead to a small systematic shift of the field-induced phase boundaries.
Interestingly, the magnetic phase diagrams reported for bulk CrPS$_4$~\cite{Peng2020,fas2025} exhibit a qualitatively similar field-induced sequence. 
Upon applying an external magnetic field, the bulk A-type antiferromagnetic ground state first evolves into a canted antiferromagnetic phase through a spin-flop transition, before eventually reaching a ferromagnetic state at higher fields~\cite{Peng2020,fas2025}. 
Our monolayer phase diagram displays a similar overall structure, but with the spin-spiral phases replacing the antiferromagnetic phases found in the bulk. 
In particular, the field drives the monolayer from the zero-field SS state through the intermediate SS$'$ state and ultimately into the ferromagnetic phase.

\begin{figure}[hbt]
    \centering
    \includegraphics[width=1\linewidth]{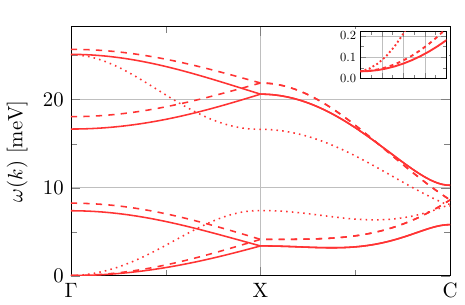}
    \caption{Linear spin-wave spectrum of a CrPS$_4$ monolayer. The high-symmetry points of the Brillouin zone are C $(\frac{1}{2},\frac{1}{2})$, $\Gamma$ $(0,0)$, and X $(\frac{1}{2},0)$. 
    Solid lines show the fully converged calculation ($N_s=14$), 
    dashed lines the spectrum obtained by truncating the exchange interactions after the fourth shell, 
    and dotted lines the spectrum obtained using the full interaction range after averaging the inequivalent first-neighbor exchange interactions (Eq.~\eqref{eq:javerage}), thereby removing the structural dimerization. 
    The inset enlarges the vicinity of the $\Gamma$ point, highlighting the effects of interaction-range truncation and dimerization on the spin stiffness and anisotropy gap.}
    \label{fig:CrPS_magnon}
\end{figure}

Finally, we computed the spin-wave spectrum of monolayer CrPS$_4$ within linear spin-wave theory using the code {\sc Magnopy}~\cite{Magnopy1,Magnopy2}.
The resulting magnon dispersion is shown in Fig.~\ref{fig:CrPS_magnon}. 
Because the optimized monolayer unit cell contains four Cr atoms, the spectrum consists of four magnon branches. 
Although neutron-scattering measurements are available only for bulk CrPS$_4$, whose crystallographic unit cell contains two Cr atoms and therefore exhibits two magnon branches~\cite{Calder2020},
the calculated spectrum reproduces the main features and energy scale of the experimental dispersion. In Ref.~\cite{Calder2020}, the mode at the $X$ point occurs at approximately 4\,meV, compared with about 3\,meV in our calculation, while the gap at the $C$ point extends from approximately 8--11\,meV experimentally and 6--10\,meV in our calculation. The overall dispersion shape is also similar throughout the Brillouin zone. In particular, the lowest magnon branch, the overall bandwidth, and the gap at the $C$ point are well reproduced, providing an independent validation of the magnetic interactions obtained from our calculations.

To assess the role of the structural dimerization, we repeated the calculation using a reduced two-atom unit cell obtained by averaging the inequivalent first-neighbor exchange interactions (dotted lines). 
This approximation noticeably deteriorates the agreement with experiment, most prominently by removing the magnon gap at the $C$ point. 
A similarly poor description is obtained when the exchange interactions are truncated after the fourth shell (dashed lines). 
Notably, in both cases the ground state remains ferromagnetic, and the degeneracy at $C$ is preserved. 
In contrast, when the full set of exchange interactions is combined with the first-neighbor dimerization, the ground state becomes spin-spiral and the degeneracy at $C$ is lifted, resulting in the finite magnon gap observed in the calculated spectrum. 
These results demonstrate that the structural dimerization and long-range exchange interactions work together to stabilize the spin-spiral state and the associated low-energy magnon structure of CrPS$_4$.

The inset of Fig.~\ref{fig:CrPS_magnon} shows an enlarged view around the $\Gamma$ point, highlighting the effects of the exchange-range truncation and lattice dimerization on the spin stiffness and the anisotropy-induced magnon gap. 
Overall, the fully \textit{ab initio} spin model reproduces the main experimental features of the available neutron-scattering spectrum, providing further support for the microscopic description developed in this work.

The CrPS$_4$ monolayer provides a clear example of the limitations of short-range spin models. 
While models truncated after three  shells predict a ferromagnetic ground state, including longer-range interactions stabilizes an exchange-driven spin spiral and reduces the critical temperature.
Furthermore, both the magnetic phase diagram and the magnon spectrum require the combined inclusion of structural dimerization and long-range exchange interactions, highlighting the importance of parameter-free exchange tensors for quantitatively predictive spin models.

\section{NiPS$_3$ Results}
\label{sec:NiPS_results}

\begin{figure}
    \centering
    \includegraphics[width=0.80\linewidth]{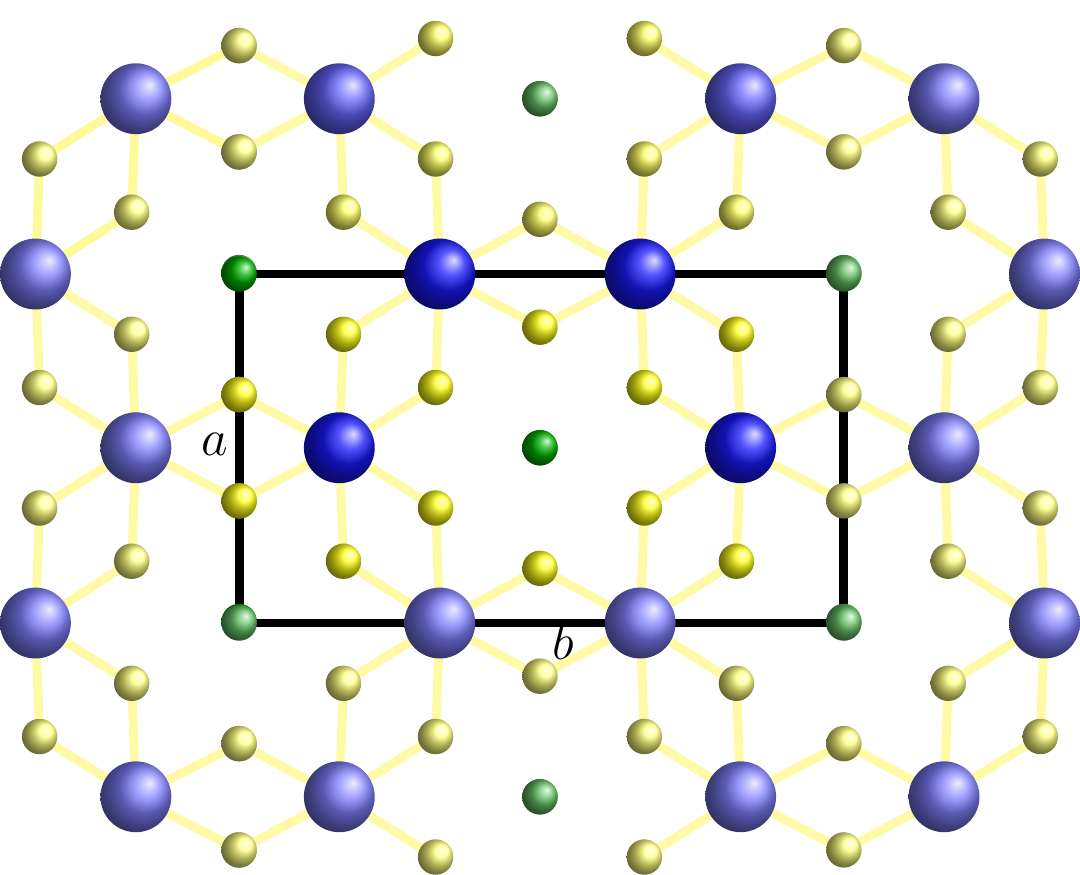}
    \caption{Optimized crystal structure of a NiPS$_3$ monolayer. 
    Blue, green, and yellow spheres denote Ni, P, and S atoms, respectively, while the yellow lines indicate the chemical bonds. The lattice constants are $a=5.91$\,\AA\ and $b=10.24$\,\AA, as indicated by the black lines.}
    \label{fig:NiPS3_structure}
\end{figure}

NiPS$_3$ crystallizes in a hexagonal lattice in which neighboring Ni atoms are connected through two S atoms, 
while two P atoms are located above and below the center of each hexagon (Fig.~\ref{fig:NiPS3_structure}). 
Its magnetic ground state consists of ferromagnetic zigzag chains coupled antiferromagnetically (see Fig~\ref{fig:NiPS_zigzags}), thereby breaking the six-fold rotational symmetry of the lattice. 
To capture this symmetry breaking, we employ the enlarged rectangular unit cell shown in Fig.~\ref{fig:NiPS3_structure}, 
which allows both the atomic positions and the magnetic structure to relax consistently. 
Throughout this work, the $x$ and $y$ axes are chosen parallel to the $a$ and $b$ lattice vectors, respectively.

Structural relaxations were first performed for several magnetic configurations using $U=0$. 
The lowest-energy state corresponds to the experimentally reported in-plane zigzag order, while the equivalent zigzag configuration with spins aligned along the $z$ direction lies only $\sim0.15$\,meV higher in energy. 
The Néel state is less stable by approximately $42$\,meV, whereas all other magnetic configurations considered (including stripy and paramagnetic states) are higher in energy by about $1$\,eV.

To determine the optimal Hubbard parameter, we compared the calculated electronic gap and spin-wave spectrum with the available experimental data~\cite{Scheie2023}.
We found that $U=4$\,eV provides the best overall agreement with the measured magnon dispersion while yielding a reasonable estimate of the electronic gap. 
All results discussed below therefore correspond to $U=4$\,eV. 
The optimized structure exhibits a small magnetostrictive distortion, with antiferromagnetic bonds being approximately $0.006$\,\AA\ shorter than ferromagnetic ones, in agreement with previous studies of van der Waals magnets~\cite{Davis2025}.
The resulting lattice constants are $a=5.91$\,\AA\ and $b=10.24$\,\AA, consistent with earlier experimental and theoretical reports~\cite{Ouvrard1985,Rao1992,Chittari2016}. 
The local magnetic moment is $1.13\,\mu_\mathrm{B}$ per Ni atom, and the calculated electronic structure exhibits an indirect band gap of $1.7$\,eV, exceeding the values reported for bulk~\cite{Foot1980} and thin-layer~\cite{Ho2021,Ran2022} NiPS$_3$ by approximately $0.2$--$0.3$\,eV.

\begin{figure}[hbt]
    \centering
    \includegraphics[width=0.66\linewidth]{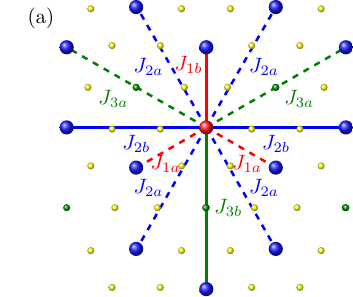}
     \includegraphics[width=0.9\linewidth]{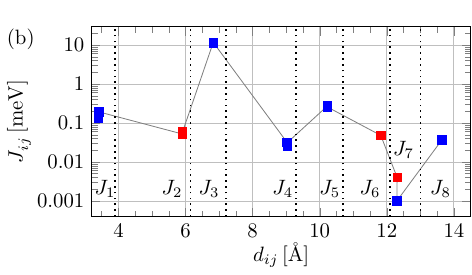}
    \caption{(a) Geometry of the first three exchange shells in a NiPS$_3$ monolayer. 
    (b) Calculated isotropic exchange interactions $J_N$ as a function of the Ni--Ni distance $d_N$ (logarithmic scale). 
    Blue and red squares denote antiferromagnetic ($J_N>0$) and ferromagnetic ($J_N<0$) interactions, respectively. The vertical dashed line show the boundaries of the exchange shells marked by $J_N$.}
    \label{fig:NiPS3_J-s}
\end{figure}

The magnetic interactions are also nearly spin-isotropic in NiPS$_3$.
The single-ion anisotropy ($A^{xx}=-28$\,$\mu$eV and $A^{yy}=-3$\,$\mu$eV) and the exchange anisotropy ($d^{xx}=-10$\,$\mu$eV and $d^{yy}=-19$\,$\mu$eV) are both in the $\mu$eV range and favor spin alignment along the $x$ direction. 
Consequently, we restrict the following discussion to the isotropic exchange interactions $J_N$.

The geometry of the first three exchange shells is shown in Fig.~\ref{fig:NiPS3_J-s}\,(a), while the calculated isotropic exchange interactions are plotted in Fig.~\ref{fig:NiPS3_J-s}\,(b).  
The magnetostrictive lattice reconstruction breaks the trigonal $C_3$ rotational symmetry of the exchange constants to monoclinic. 
The relative difference between the inequivalent couplings $J_a$ and $J_b$ decreases rapidly with distance, amounting to 43\%, 15\%, and 1.5\% for the first, second, and third shells, respectively, and becoming negligible for more distant neighbors.

In agreement with previous theoretical and experimental studies~\cite{lancon2018,Wildes2022,orlando2026,Scheie2023}, we find that the dominant exchange interaction is the antiferromagnetic third-neighbor coupling $J_3$. 
In contrast to earlier fitted spin models, however, our \emph{ab initio} calculations predict a weak antiferromagnetic first-neighbor interaction and an almost negligible ferromagnetic second-neighbor interaction. 
With these exchange parameters, a model restricted to the first three  shells stabilizes a Néel antiferromagnetic ground state rather than the experimentally observed zigzag order. 
Within such a three-shell description, a sufficiently strong ferromagnetic first-neighbor interaction with would be required to introduce the magnetic frustration needed to stabilize the zigzag phase. 
Our calculations show instead that the missing frustration originates from the fifth-neighbor interaction, which is antiferromagnetic and satisfies $J_5\approx2J_1$. Including this interaction stabilizes the zigzag ground state, demonstrating that exchange interactions beyond the third shell are essential for a quantitatively predictive description of monolayer NiPS$_3$.

\begin{figure}[hbt]
    \centering
            \includegraphics[width=0.99\linewidth]{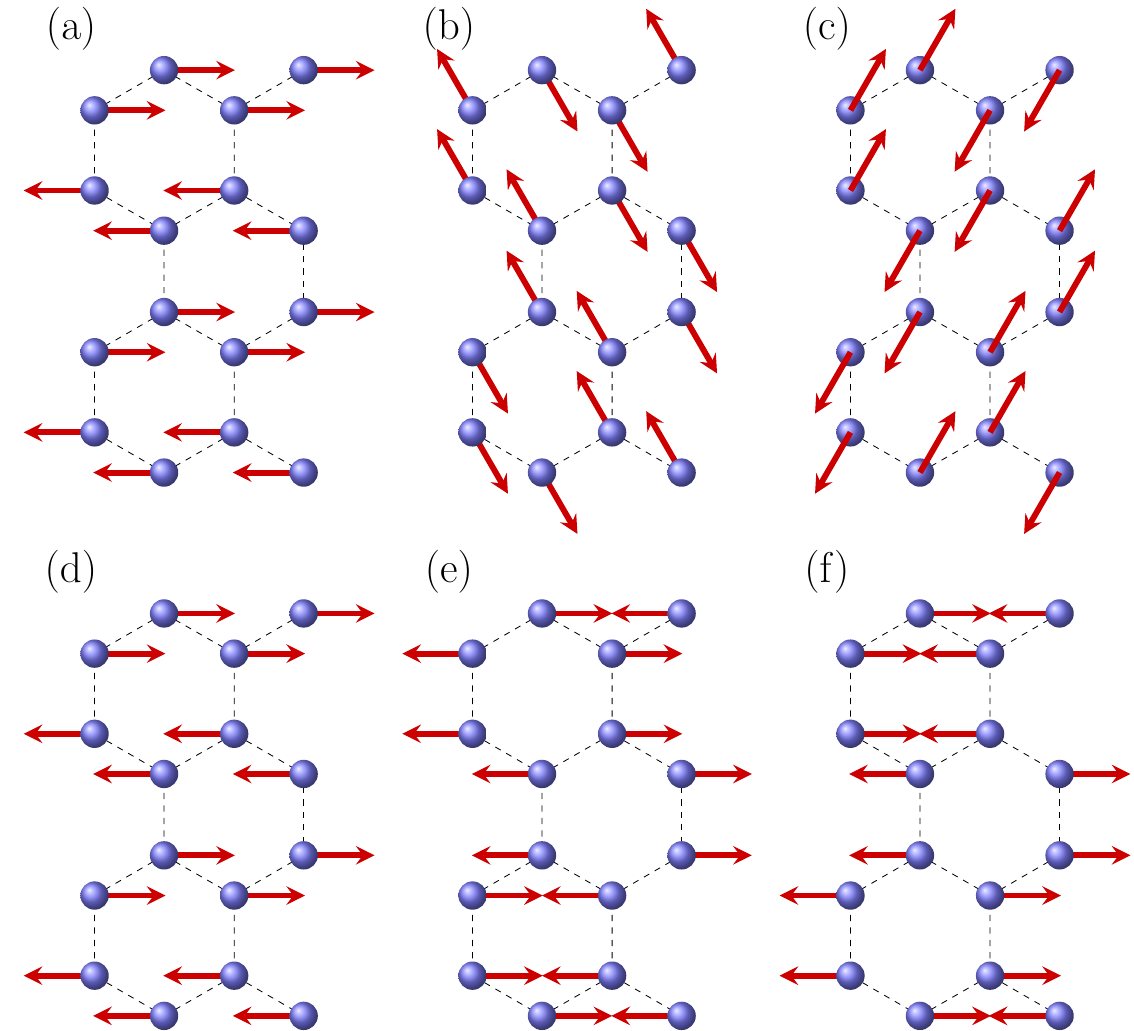}
            \caption{Symmetry-equivalent zigzag magnetic configurations in the ideal hexagonal lattice ((a)--(c)) and the corresponding ground states obtained after including the anisotropy induced by magnetostriction ((d)--(f)).}
    \label{fig:NiPS_zigzags}
\end{figure}

Monte Carlo simulations reproduce the experimentally observed zigzag antiferromagnetic ground state of monolayer NiPS$_3$. 
The magnetic moments are oriented predominantly along the $x$ direction, with a small canting of approximately $11^\circ$ towards the $z$ axis.
In the absence of magnetostriction, the hexagonal lattice possesses $C_{3}$ symmetry, giving rise to three symmetry-equivalent zigzag domains with propagation directions at $0^\circ$ and $\pm120^\circ$, as illustrated in Fig.~\ref{fig:NiPS_zigzags}\,(a)--(c). 
Spin anisotropy selects the $x$ direction as the easy axis, while the weak anisotropy of the exchange interactions induced by magnetostriction slightly lifts the degeneracy between the three propagation directions without changing the preferred spin orientation, leading to the configurations shown in Fig.~\ref{fig:NiPS_zigzags}\,(d)--(f). 
The resulting energy splitting is extremely small: the two domains with propagation vectors at $\pm120^\circ$ differ by less than $0.2\,\mu$eV per Ni atom, whereas the domain propagating along the $0^\circ$ direction is lower in energy by approximately $15\,\mu$eV. 
It is important to note that the exchange interactions were extracted from a single self-consistent {\sc SIESTA} calculation for the relaxed $0^\circ$ zigzag structure. 
The energies of the remaining zigzag domains were then evaluated using the same spin Hamiltonian, without allowing for further lattice relaxation or recalculating the exchange interactions. 
Consequently, the small energy differences originate solely from the anisotropy already encoded in the exchange interactions of the relaxed $0^\circ$ structure. 
The six nearly degenerate magnetic states—three propagation directions combined with two opposite spin orientations—therefore constitute an almost ideal realization of a weakly perturbed six-state clock model~\cite{sixclock}.

\begin{figure}[hbt]
    \centering
       \includegraphics[width=0.9\linewidth]{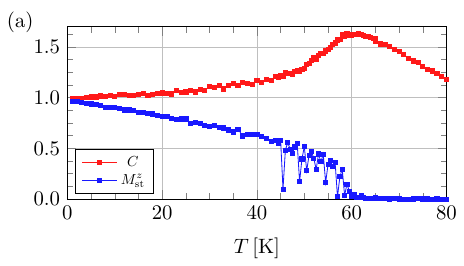}
        \includegraphics[width=0.9\linewidth]{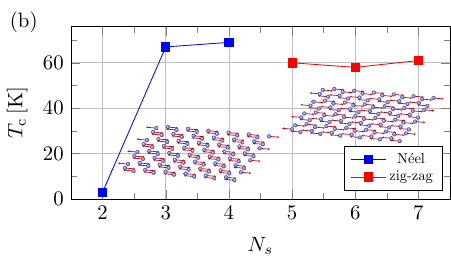}
    \caption{(a) Heat capacity (red) and staggered magnetisation  (blue) (Eq.~\eqref{eq:stag}) of NiPS$_3$ as a function of temperature, calculated with full set of magnetic interactions.
    (b) Critical temperature $T_\mathrm{c}$ of a NiPS$_3$ monolayer as a function of the number of exchange shells $N_s$ included in the spin Hamiltonian. Blue and red squares correspond to Néel and zigzag antiferromagnetic ground states, respectively. The change in color marks the transition between the two magnetic ground states as the interaction range is extended.}
    \label{fig:NiPS3_Tc}
\end{figure}

To determine the critical temperature, we analyzed the heat capacity, magnetic susceptibility, and staggered magnetization as functions of temperature. The latter is defined as
\begin{align}
M^z_\mathrm{st}=M^z_1+M^z_2-M^z_3-M^z_4\,,
\label{eq:stag}
\end{align}
where $M^z_i$, with $i=1,2,3,4$, denotes the $z$-component of the magnetization of the four magnetic atoms in the unit cell.
The peak in the heat capacity and the disappearance of the zigzag order consistently identify the transition to the paramagnetic phase at $T_\mathrm{c}=61$\,K (Fig.~\ref{fig:NiPS3_Tc}(a)). 
This value is the half of the value 120\,K what was measured in Ref.~\cite{Cheon2025}.
The difference between the experiment and the calculations can be originated to the fact, that a classical simulation overestimates the thermal disorder and systematically underestimates the ordering temperature~\cite{evans2015,woo2015}. Mendili and Zhitomirsky \cite{mendili2026} argued that classical exchange and temperature scales must be multiplied by $S(S+1)/S^2$, which, for an $S=1$ system such as NiPS$_3$ mean a factor of 2.
We then investigated the dependence of the magnetic ground state and critical temperature on the number of exchange shells $N_s$ included in the spin Hamiltonian (Fig.~\ref{fig:NiPS3_Tc}(b)). 
Restricting the model to the first three shells yields a Néel antiferromagnetic ground state with $T_\mathrm{c}=68$\,K.
Including the fourth shell leaves both the magnetic structure and the transition temperature essentially unchanged. 
In contrast, the fifth-neighbor interaction stabilizes the experimentally observed zigzag phase, while the critical temperature remains close to $61$\,K. 
Extending the interaction range beyond the fifth shell produces no further changes in either the magnetic ground state or the critical temperature, demonstrating that a five-shell model is both necessary and sufficient to accurately describe monolayer NiPS$_3$ when the exchange interactions are determined fully \emph{ab initio}.

We investigated the effect of external magnetic fields applied both parallel and perpendicular to the monolayer plane, with strengths up to 10\,T. 
The zigzag antiferromagnetic ground state remains remarkably robust throughout this range.
For both field orientations, the spin moments rotate by only about $0.2^\circ$ at 2\,T, while no qualitative changes in the magnetic structure are observed up to the largest field considered. 
Likewise, the critical temperature remains essentially unchanged over the entire field range. 
These results demonstrate that the zigzag phase of monolayer NiPS$_3$ is remarkably insensitive to experimentally accessible magnetic fields.

\begin{figure}[hbt]
    \centering
    \includegraphics[width=0.9\linewidth]{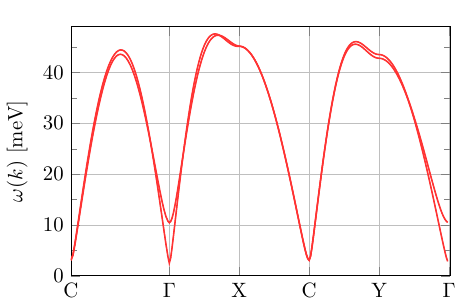}
    \caption{Linear spin-wave spectrum of a NiPS$_3$ monolayer. The high-symmetry points of the Brillouin zone are C $(\frac{1}{2},\frac{1}{2})$, $\Gamma$ $(0,0)$, X $(\frac{1}{2},0)$, and Y $(0,\frac{1}{2})$.}
    \label{fig:NiPS_magnom}
\end{figure}

We finally computed the linear spin-wave spectrum of monolayer NiPS$_3$ using {\sc Magnopy} 
and compared it with the spectrum obtained from neutron-scattering measurements of bulk NiPS$_3$ in Fig.~5 of Ref.~\cite{Scheie2023}. 
The Hubbard parameter $U$ has a strong influence on the magnon bandwidth: calculations with $U=0$ yield a bandwidth of approximately $150\,$meV, significantly larger than the $\sim40$\,meV measured experimentally~\cite{Scheie2023}. 
Increasing $U$ progressively reduces the bandwidth, and $U=4$\,eV provides the best overall agreement with the experimental dispersion.
The corresponding spin-wave spectrum is shown in Fig.~\ref{fig:NiPS_magnom}. 
Some discrepancies remain, particularly in the magnon gaps, which are overestimated in our calculations. 
These differences may originate from effects absent in the monolayer model, such as interlayer exchange interactions and magnetoelastic couplings, both of which are expected to influence the magnetic excitations of bulk NiPS$_3$. 
This interpretation is consistent with the experimentally observed differences between monolayer and multilayer NiPS$_3$~\cite{kim2019,Hu2023}.

Taken together with the results for CrPS$_4$, the present analysis demonstrates that reliable spin models for transition-metal thiophosphates require exchange interactions extending well beyond the third shell. 
While the specific mechanisms differ between the two compounds, long-range interactions are essential in both cases to correctly describe the magnetic ground state and to achieve quantitative agreement with experiment.

\section{Conclusions}
\label{sec:conc}

We have presented a comprehensive first-principles study of the magnetic properties of monolayer CrPS$_4$ and NiPS$_3$, combining density functional theory, the LKAG formalism for extracting tensorial exchange interactions, Monte Carlo simulations, and linear spin-wave theory calculations. 
This approach enables the construction of predictive spin models directly from \textit{ab initio} calculations, without relying on total-energy fitting, while systematically assessing the role of long-range exchange interactions.

Our results demonstrate that exchange interactions extending well beyond the third  shell are essential for a quantitatively predictive description of both materials. 
In CrPS$_4$, long-range interactions destabilize the previously predicted ferromagnetic ground state in favor of a spin-spiral phase and reduce the critical temperature to approximately 21\,K, in excellent agreement with experiment. 
The resulting magnetic phase diagram exhibits a rich sequence of collinear and non-collinear phases that can be tuned by temperature and external magnetic fields. 
In NiPS$_3$, the fifth-neighbor exchange interaction provides the frustration required to stabilize the experimentally observed zigzag antiferromagnetic order, while the calculated spin-wave spectrum reproduces the main features of the available neutron-scattering data.

Taken together, these results demonstrate that the commonly adopted three-shell Heisenberg models are insufficient to describe transition-metal thiophosphate monolayers quantitatively. 
Instead, reliable microscopic spin models require long-range exchange interactions obtained directly from first principles. 
We expect that the methodology presented here will be broadly applicable to the accurate description and prediction of magnetic phenomena in two-dimensional van der Waals materials.


\begin{acknowledgments}
J. F. acknowledges discussions with G. Martínez-Carracedo, A. Rybakov and R. den Teuling.
The authors are grateful for all discussion and financial support.
B. N., A. G. F. and J. F. have been funded by MCIN/AEI/10.13039/501100011033/FEDER, UE via project PID2022-137078NB-100 and PID2025-169100NB-I00, by the TRILMAX Horizon Europe consortium (Grant No. 101159646), and by Agencia SEKUENS (Asturias) under grant UONANO 
IDE/2024/000678 with the support of FEDER funds.
R. D. and Y. M. B. acknowledge support by the Dutch Research Council (NWO) under the project "Ronde Open Competitie ENW 
pakket 21-3" (file number OCENW.M.21.215) which is (partly) financed by the Dutch Research Council (NWO).
A. B.-S., H. S. J. vdZ. and Y. M. B. acknowledge support by the Dutch Research Council (NWO) under the project “Ronde Open Competitie XL” (file number OCENW.XL21.XL21.058).
\end{acknowledgments}

\appendix

\bibliography{citations}

\end{document}